\documentclass[11pt, letterpaper]{article}

\usepackage[margin=1in]{geometry}
\usepackage{times}
\usepackage{hyperref}
\usepackage{graphicx}
\usepackage{url}
\usepackage{amsmath}
\usepackage{amssymb}
\usepackage{enumitem}
\usepackage{booktabs} 
\usepackage{caption}

\usepackage{algorithm}
\usepackage{algpseudocode}

\usepackage{pgfplots}
\usepackage{tikz}
\usetikzlibrary{positioning, arrows.meta, shapes}
\pgfplotsset{compat=1.18}

\title{A Chunked-Object Pattern for Multi-Region Large Payload Storage \\ in Managed NoSQL Databases}

\author{
  Manideep Reddy Chinthareddy \\
  \texttt{Centerville, VA, USA }\\
  \texttt{chmanideepreddy@gmail.com}
}

\date{December 2025}

\begin{document}

\maketitle

\begin{abstract}
Many managed key--value and NoSQL databases---such as Amazon DynamoDB, Azure Cosmos DB, and Google Cloud Firestore---enforce strict maximum item sizes (e.g., 400\,KB in DynamoDB). This constraint imposes significant architectural challenges for applications requiring low-latency, \emph{multi-region} access to objects that exceed these limits. The standard industry recommendation is to offload payloads to object storage (e.g., Amazon S3) while retaining a pointer in the database. While cost-efficient, this ``pointer pattern'' introduces network overhead and exposes applications to non-deterministic replication lag between the database and the object store, creating race conditions in active-active architectures.

This paper presents a ``chunked-object'' pattern that persists large logical entities as sets of ordered chunks within the database itself. We precisely define the pattern and provide a reference implementation using Amazon DynamoDB Global Tables. The design generalizes to any key--value store with per-item size limits and multi-region replication. We evaluate the approach using telemetry from a production system processing over 200,000 transactions per hour. Results demonstrate that the chunked-object pattern eliminates cross-system replication lag hazards and reduces $p99$ cross-region time-to-consistency for 1\,MB payloads by keeping data and metadata within a single consistency domain.
\end{abstract}

\section{Introduction}
Managed NoSQL databases are widely adopted for latency-sensitive, high-throughput applications due to their horizontal scalability and managed multi-region replication. However, systems such as Amazon DynamoDB~\cite{dynamo}, Azure Cosmos DB, and Google Cloud Firestore impose strict limits on item size (400\,KB, 2\,MB, and 1\,MB, respectively). Applications managing moderately sized payloads---such as aggregated analytics snapshots or large JSON documents---must often offload data to separate object storage services.

The prevailing design pattern involves storing the payload in object storage (e.g., Amazon S3) and referencing it via a pointer in the database~\cite{aws-dynamodb}. While effective for archival data, this approach introduces heterogeneous consistency models. In multi-region ``active-active'' architectures, the database and object store replicate asynchronously and independently. This creates a hazard where a secondary region may receive the database pointer before the object storage payload has replicated, leading to ``dangling pointer'' errors.

In this paper, we propose and evaluate a database-centric \emph{Chunked-Object Pattern}. By splitting logical entities into fragments that fit within the database's item limit, we enforce atomic visibility of the payload and metadata within a single replication stream.

Our contributions are:
\begin{itemize}[noitemsep]
  \item A precise definition of the chunked-object pattern for overcoming item size limits in distributed NoSQL stores.
  \item A multi-cloud analysis of replication behavior, highlighting the risks of hybrid storage (Database + Object Store) in active-active topologies.
  \item An empirical evaluation based on production telemetry comparing the proposed pattern against the traditional S3-pointer approach, demonstrating the elimination of cross-region race conditions.
\end{itemize}

\section{Background and Motivation}

\subsection{The Multi-Region Replication Gap}
Managed databases typically offer optimized replication for small items. DynamoDB Global Tables and Cosmos DB, for instance, utilize log-based replication that propagates changes in sub-seconds to seconds. Conversely, object storage replication (e.g., S3 Cross-Region Replication) is throughput-optimized rather than latency-optimized.

In a typical active-active design, a write request may land in Region~A, followed immediately by a read or update in Region~B. If the application relies on the ``Pointer Pattern,'' correctness depends on the convergence of two distinct distributed systems. Since no major cloud provider offers a cross-region replication latency SLA for object storage, the ``Time-to-Consistency'' is non-deterministic, often trailing the database by seconds or minutes under load.

\subsection{Motivating Case Study}
This research is motivated by a production financial services system requiring multi-region availability for request payloads ranging from 400\,KB to 2\,MB. The system processes approximately 55 transactions per second. Initial stress testing of the standard S3-Pointer pattern revealed that while database metadata replicated in $<2$ seconds, S3 payloads frequently lagged by 15--30 seconds. This discrepancy necessitated complex application-side retry logic and exponential backoff, violating the system's latency Service Level Objectives (SLOs).

\section{The Chunked-Object Pattern}

\subsection{Data Model}
We represent a logical entity $E$ with identifier $id$ as a collection of database items sharing a Partition Key (PK).
\begin{itemize}[noitemsep]
  \item \textbf{Metadata Record:} Stores versioning, checksums, and the total chunk count.
  \item \textbf{Chunk Records:} Store binary fragments of the payload.
\end{itemize}

By utilizing the Sort Key (SK) for ordering (e.g., \texttt{CHUNK\#001}, \texttt{CHUNK\#002}), the database allows efficient retrieval of all fragments via a single range query.

\subsection{Write Protocol}
To ensure consistency, we utilize an optimistic locking strategy. The write operation is formalized in Algorithm~\ref{alg:chunked_write}.

\begin{algorithm}
\caption{Chunked-Object Write Protocol}
\label{alg:chunked_write}
\begin{algorithmic}[1]
\Require Entity Identifier $ID$, Binary Payload $P$, Max Chunk Size $C_{max}$
\Ensure Payload persisted to database

\State $V_{new} \gets \text{GenerateMonotonicVersion}()$
\State $Chunks \gets \text{Split}(P, C_{max})$
\State $WriteBatch \gets \emptyset$

\Statex \Comment{\textbf{Step 1: Prepare Chunk Items}}
\For{$i \gets 0$ \textbf{to} $\text{Length}(Chunks) - 1$}
    \State $Item_{chunk} \gets \{ \text{PK}: ID, \text{SK}: \text{CHUNK}\#i, \text{Ver}: V_{new}, \text{Data}: Chunks[i] \}$
    \State $\text{Add } Item_{chunk} \text{ to } WriteBatch$
\EndFor

\Statex \Comment{\textbf{Step 2: Prepare Metadata Item}}
\State $Item_{meta} \gets \{ \text{PK}: ID, \text{SK}: \text{META}, \text{Ver}: V_{new}, \text{Count}: \text{Length}(Chunks) \}$
\State $\text{Add } Item_{meta} \text{ to } WriteBatch$

\Statex \Comment{\textbf{Step 3: Execute Write}}
\If{$\text{Size}(WriteBatch) \leq \text{TransactionLimit}$}
    \State \textbf{TransactWriteItems}($WriteBatch$)
\Else
    \State \Comment{Fallback for very large payloads}
    \State \textbf{TwoPhaseWrite}($WriteBatch$, $Item_{meta}$)
\EndIf
\end{algorithmic}
\end{algorithm}

As outlined in Algorithm~\ref{alg:chunked_write}, inclusion of the metadata item in the transactional batch serves as a commit barrier. Readers query the metadata first; if the metadata for version $V$ exists, the database guarantees (via transactional atomicity) that the corresponding chunks also exist.

For payloads that exceed the transactional item limit, \textbf{TwoPhaseWrite} first writes chunk records with a provisional status (e.g., \texttt{status = WRITING}) and only marks the metadata record as committed (e.g., \texttt{status = COMMITTED}) once all chunks have been successfully persisted. Readers ignore versions that are not yet in the COMMITTED state, ensuring that partially written entities are never exposed.

\subsection{Read Path and Reassembly}
The read operation performs a query for \texttt{PK = ID}. This efficiently retrieves the metadata and all chunk items in a single network round-trip (provided the result set fits within the database's page size). The application validates the chunk count against the metadata before concatenating the binary data. If the number of retrieved chunks does not match the metadata count, or if checksums do not validate, the reader can fall back to a previous version or retry.

\subsection{Applicability to Cosmos DB and Firestore}
Although the reference implementation targets DynamoDB, the pattern maps naturally to other managed NoSQL systems:

\begin{itemize}[noitemsep]
  \item In \textbf{Cosmos DB}, the partition key plays the role of PK, while a synthetic sort attribute (e.g., \texttt{type + chunkIndex}) can act as SK. The chunked-object pattern can take advantage of Cosmos DB's transactional batch support within a partition and its tunable consistency levels (e.g., Bounded Staleness) to bound replica lag.
  \item In \textbf{Firestore}, chunks can be modeled as documents in a subcollection under a parent document that stores metadata (version, count). Firestore's strong global consistency simplifies reasoning about versioning and conflict resolution, at the cost of different performance and pricing trade-offs.
\end{itemize}

In all three systems, the core idea of a metadata record plus ordered chunk records remains unchanged.

\section{Multi-Region Behavior Analysis}

We analyze the replication characteristics of the three major cloud providers in Table~\ref{tab:nosql-mr}. A consistent finding is that while NoSQL stores offer tunable consistency or low-latency replication, object stores prioritize durability and throughput over replication speed.

\begin{table}[h]
\centering
\small
\caption{Multi-Region Capabilities of Managed NoSQL Databases}
\label{tab:nosql-mr}
\begin{tabular}{@{}llll@{}}
\toprule
\textbf{Feature} & \textbf{DynamoDB} & \textbf{Cosmos DB} & \textbf{Firestore} \\ \midrule
Architecture & Active-Active & Active-Active & Multi-Region Instance \\
Replication Lag & $<1$s (Eventual) & 10--100ms & Sync/Near-Sync \\
Consistency Model & Eventual & Tunable* & Strong \\
Active Writes & Multi-Master & Multi-Master & Yes (multi-region instance) \\
Object Store Peer & S3 & Blob Storage & Cloud Storage \\
Obj.\ Store Lag SLA & None & None & None \\
Max Item Size & 400\,KB & 2\,MB & 1\,MB \\ \bottomrule
\multicolumn{4}{l}{\footnotesize *Includes Strong, Bounded Staleness, Session, Consistent Prefix, and Eventual.}
\end{tabular}
\end{table}

The key observation from Table~\ref{tab:nosql-mr} is the architectural mismatch. While databases like Cosmos DB and DynamoDB enable sub-second cross-region propagation (via log-shipping), the associated object storage services (S3, Blob, Cloud Storage) operate on asynchronous, throughput-optimized queues. This creates a ``replication gap'' where metadata exists in the destination region before the payload arrives.

\section{Empirical Evaluation}
To validate the pattern, we analyzed telemetry data from the production environment described in Section 2.2 over a 6-month period. The system operates two distinct microservices—one utilizing the legacy S3-Pointer pattern and another utilizing the Chunked-Object pattern—processing a combined peak load of approximately 200,000 transactions per hour.

\subsection{Cross-Region Replication Latency}
\begin{figure}[h]
    \centering
    \begin{tikzpicture}
        \begin{axis}[
            ybar,
            title={Cross-Region Replication Latency (1MB Payload)},
            ylabel={Latency (seconds)},
            symbolic x coords={p50, p95, p99},
            xtick=data,
            nodes near coords,
            legend style={at={(0.5,-0.15)}, anchor=north, legend columns=-1},
            ymin=0, ymax=35,
            grid=major,
            width=8cm,
            height=6cm
        ]
        \addplot coordinates {(p50, 0.4) (p95, 0.9) (p99, 1.8)};
        
        \addplot coordinates {(p50, 1.2) (p95, 4.5) (p99, 28.5)};
        
        \legend{Chunked-Object, S3-Pointer}
        \end{axis}
    \end{tikzpicture}
    \caption{Comparison of replication latency. The Chunked-Object pattern (Blue) maintains tight consistency, whereas the S3-Pointer pattern (Red) exhibits significant tail latency ($p99 \approx 28.5s$) due to asynchronous object replication.}
    \label{fig:replication_latency}
\end{figure}
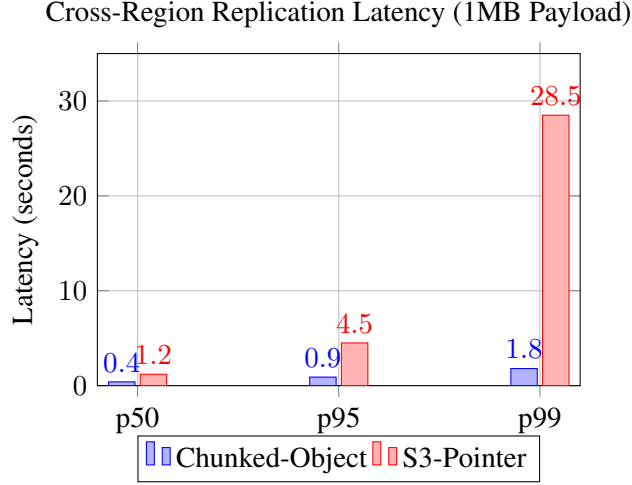

We compared the ``Time-to-Consistency'' between the S3-Pointer implementation and the Chunked-Object implementation. Figure~\ref{fig:replication_latency} illustrates the replication lag for a 1\,MB payload.

The S3-Pointer pattern exhibited a ``long tail'' behavior. While median performance was acceptable, the $p99$ lag extended to nearly 30 seconds during periods of high throughput.

\textbf{Worst-Case Analysis:} During peak load windows ($>200$k transactions/hour), we observed significant divergence in worst-case behavior. The S3-based microservice experienced replication delays exceeding \textbf{180 seconds (3 minutes)} for multi-file transactions, resulting in persistent \texttt{404 Not Found} errors for cross-region consumers. In contrast, the DynamoDB-backed implementation, while also experiencing lag increases due to congestion, remained strictly bounded within \textbf{5 seconds}. This demonstrates that the chunked pattern offers deterministic upper bounds on consistency even under saturation, whereas object storage replication remains unbounded.

\subsection{The ``Dangling Pointer'' Race Condition}
The disparity in replication speeds illustrated above introduces a critical race condition in active-active architectures. This failure mode, which we term the ``Dangling Pointer Hazard,'' occurs because the metadata channel is faster than the payload channel.

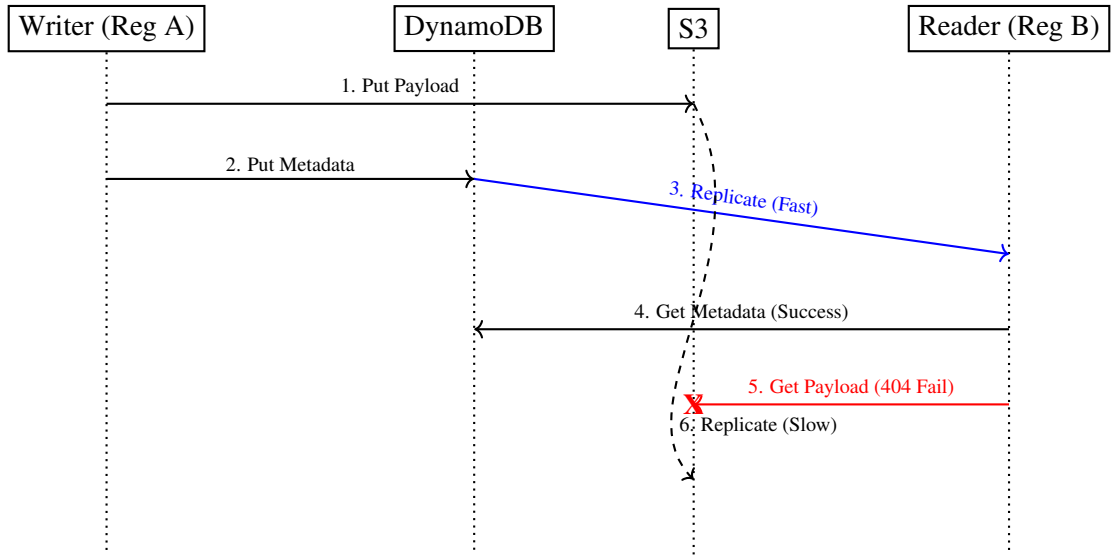
\begin{figure}[h]
    \centering
    \begin{tikzpicture}[node distance=1.5cm, auto, thick]
        \node (Client) [draw, rectangle, minimum width=1.5cm] {Writer (Reg A)};
        \node (DB) [draw, rectangle, right=2.5cm of Client] {DynamoDB};
        \node (S3) [draw, rectangle, right=1.5cm of DB] {S3};
        \node (Reader) [draw, rectangle, right=2.5cm of S3] {Reader (Reg B)};

        \draw[dotted] (Client) -- ++(0,-7);
        \draw[dotted] (DB) -- ++(0,-7);
        \draw[dotted] (S3) -- ++(0,-7);
        \draw[dotted] (Reader) -- ++(0,-7);

        \coordinate (t0) at (0,-1);
        \coordinate (t1) at (0,-2);
        \coordinate (t2) at (0,-3);
        \coordinate (t3) at (0,-4);
        \coordinate (t4) at (0,-5);
        \coordinate (t5) at (0,-6);

        \draw[->] (Client |- t0) -- node[above, scale=0.7] {1. Put Payload} (S3 |- t0);
        \draw[->] (Client |- t1) -- node[above, scale=0.7] {2. Put Metadata} (DB |- t1);

        \draw[->, color=blue] (DB |- t1) -- node[above, sloped, scale=0.7] {3. Replicate (Fast)} (Reader |- t2);
        
        \draw[->] (Reader |- t3) -- node[above, scale=0.7] {4. Get Metadata (Success)} (DB |- t3);
        \draw[->, color=red] (Reader |- t4) -- node[above, scale=0.7] {5. Get Payload (404 Fail)} (S3 |- t4);
        \node[color=red] at (S3 |- t4) {\textbf{X}};

        \draw[->, dashed] (S3 |- t0) .. controls +(1,-2) and +(-1,1) .. node[right, scale=0.7, pos=0.8] {6. Replicate (Slow)} (S3 |- t5);
        
    \end{tikzpicture}
    \caption{Sequence Diagram of the Dangling Pointer Hazard. The Reader in Region B successfully reads the metadata (Step 4) but fails to read the payload (Step 5) because S3 replication (Step 6) lags behind DynamoDB replication.}
    \label{fig:sequence_diagram}
\end{figure}

Figure~\ref{fig:sequence_diagram} visualizes the race condition. The Reader in Region B receives the metadata almost immediately ($t_1$) but attempts to fetch the payload ($t_2$) before S3 replication completes ($t_3$).

Table~\ref{tab:failure_rates} quantifies this failure rate. We measured the probability of a \texttt{404} error when performing a read in the secondary region immediately following the appearance of the metadata.

\begin{table}[h]
    \centering
    \caption{Secondary Region Read Failures (Dangling Pointer Errors)}
    \label{tab:failure_rates}
    \begin{tabular}{@{}lcc@{}}
        \toprule
        \textbf{Metric} & \textbf{S3-Pointer} & \textbf{Chunked-Object} \\ \midrule
        Replication Model & Decoupled & Unified \\
        Avg. Lag Delta & $4.2$s & $0.0$s \\
        $p99$ Lag Delta & $28.5$s & $<0.1$s \\
        \textbf{404 Error Rate} & \textbf{12.4\%} & \textbf{$<$ 0.01\%} \\ \bottomrule
    \end{tabular}
\end{table}

By migrating to the Chunked-Object pattern, the system reduced the cross-region 404 error rate from 12.4\% to near zero, validating the hypothesis that unifying the replication stream eliminates the dangling pointer hazard.

\section{Related Work}
The concept of splitting large objects is well-established in distributed systems. The Google File System (GFS) utilizes fixed-size chunks to manage large files across distributed nodes~\cite{gfs}. Similarly, Log-Structured Merge-trees (LSM), the underlying storage structure for DynamoDB and Cassandra, manage write amplification by segmenting data~\cite{lsm}.

The challenge of synchronizing large state across geo-distributed nodes is not unique to databases. Recent work in Edge AI, such as DisCEdge~\cite{discedge}, utilizes similar tokenization and replication strategies to ensure user context travels consistently across regions, validating the efficacy of granular data management in high-latency environments. Our work adapts these fundamental principles to the application layer of managed NoSQL databases. While cloud providers advocate for the ``Claim Check'' pattern (offloading to S3)~\cite{aws-s3}, this paper demonstrates that for active-active multi-region architectures, application-side chunking provides superior consistency guarantees by keeping both metadata and payload within a single replication and consistency domain.

\section{Conclusion}
We presented a chunked-object pattern for storing large logical entities in managed NoSQL databases. By retaining data within the database's replication boundary, the pattern eliminates cross-system race conditions inherent to hybrid storage designs. Production data confirms that this approach significantly improves $p99$ cross-region consistency latency and availability in active-active topologies.

\bibliographystyle{abbrv}

\end{document}